# A lower limit of Δz > 0.06 for the duration of the reionization epoch

Judd D. Bowman[1]* & Alan E. E. Rogers[2]*

[1]Arizona State University, **School of Earth and Space Exploration,** Tempe, Arizona 85287, USA.

[2]Massachusetts Institute of Technology, Haystack Observatory, Westford, Massachusetts, 01886 USA.

*These authors contributed equally to this work.

**Observations of the 21-centimetre line of atomic hydrogen in the early Universe directly probe the history of the reionization of the gas between galaxies[1]. The observations are challenging, though, because of the low expected signal strength (~10 mK), and contamination by strong (>100 K) foreground synchrotron emission in the Milky Way and extragalactic continuum sources[2]. If reionization happened rapidly, there should be a characteristic signature[2–4] visible against the smooth foreground in an all-sky spectrum. Here we report an all-sky spectrum between 100 and 200 MHz, corresponding to the redshift range 6 < z < 13 for the 21-centimetre line. The data exclude a rapid reionization timescale of Δz < 0.06 at the 95% confidence level.**

The observable differential brightness temperature caused by the redshifted 21-cm line from a volume of hydrogen gas in the intergalactic medium can be calculated from basic principles[5] and is:

$$\delta T_{21}(\boldsymbol{\theta}, z) \approx 27\,(1+\delta)\, x_{\mathrm{H\,I}} \left(1 - \frac{T_\gamma}{T_\mathrm{S}}\right) \left(\frac{1+z}{10}\right)^{1/2} \mathrm{mK} \qquad (1)$$

where $\theta$ is the position on the sky, $z$ is the redshift of the gas, the factor of 27 mK comes from cosmological factors, $\delta$ is the local matter overdensity of the gas, $x_{\mathrm{H\,I}}$ is the neutral fraction of the gas, $T_\mathrm{S}$ is the 'spin' temperature that describes the relative population of the ground and excited states of the hyperfine transition, and $T_\gamma$ is the temperature of the cosmic microwave background (CMB) radiation. The intensity of the 21-cm emission or absorption relative to the CMB has a strong dependence on the neutral fraction and the spin temperature, both of which are sensitive to the ultraviolet and X-ray radiation[2,3] from the formation of luminous sources, including early stars, galaxies and black holes.

During the reionization epoch, after the heating of the intergalactic medium, the spin temperature is expected to be much larger than the CMB temperature[6] ($T_S \gg T_\gamma$) and the 21-cm perturbations will be seen in emission against the CMB and dominated by variations in the neutral fraction. Under the additional assumption that the local neutral fraction of the gas is not correlated to the local matter overdensity, the angle-averaged form of equation (1) can be reduced to:

$$\langle \delta T_{21}(\boldsymbol{\theta}, z) \rangle_\theta \equiv \delta \overline{T}_{21}(z) \approx 27 \left( \frac{1+z}{10} \right)^{1/2} \overline{x}_{H\,I}(z) \ \ \text{mK} \qquad (2)$$

where we have explicitly written the redshift dependence of the mean neutral fraction as $\overline{x}_{H\,I}(z)$. This 'global' 21-cm signal should be observable in a measurement of the all-sky low-frequency radio spectrum through the mapping from redshift to frequency for the 21-cm line, according to: $\nu = 1{,}420/(1 + z)$ MHz. Here, 1,420 MHz is the rest-frame frequency of the 21-cm line and the redshift range appropriate for the reionization epoch is $z > 6$. Fixing the overall amplitude factor in equation (2) through the choice of a particular cosmological model, for example, the WMAP7 best-fit $\Lambda$CDM model[7,8], the global 21-cm signal becomes a direct probe of the evolution of the mean neutral fraction of hydrogen gas in the intergalactic medium during the reionization epoch.

The global 21-cm signal is challenging to observe in practice because the low-frequency radio sky is dominated by intense synchrotron emission from our own Galaxy that is more than four orders of magnitude brighter than the signal. Galactic and extragalactic free-free emission provide additional foregrounds[7], as do numerous radio point sources from active galactic nuclei, radio galaxies and local Galactic objects. Radio-frequency interference from television, FM (frequency modulated) radio, low-Earth-orbit satellites, and other telecommunications transmitters are prolific and can be eight or ten orders of magnitude brighter than the astrophysical signal, even in geographically remote areas.

We deployed a custom-built, high-dynamic-range broadband radio spectrometer, called EDGES[9,10], at the Murchison Radio-astronomy Observatory in Western Australia, to measure the radio spectrum between 100 and 200 MHz. The instrument observed continuously for three months with low duty cycle and yielded the spectrum shown in Fig. 1—an average over nearly the entire southern celestial hemisphere. To overcome the foreground signal, we relied on the expectation that all of the foregrounds have smooth continuum spectra that are well modelled as simple power laws in frequency[2,11] (or redshift), and that any uncertainties in the calibration of the spectrometer could introduce only smooth spectral deviations into the measurement. We designed the instrument to accomplish this goal by shortening the electrical path length between the antenna and the internal calibration source to less than a wavelength and by using an electrically compact dipole antenna. Additional details on the design of the instrument are presented in the Supplementary Information.

The observed spectrum was fitted by a model that consists of a 21-cm signal term and a polynomial term that accounts for the foregrounds and calibration uncertainties according to: $T_{obs}(z) = \delta \bar{T}_{21}(z) + T_F(z)$, where $T_F(z) = \sum_{n=0}^{m} a_n z^n$ is the foreground term. The 21-cm term is given by equation (2) with $\bar{x}_{HI}(z) = \frac{1}{2}\tanh[(z-z_r)/\Delta z]$, following the recent convention of the WMAP7 analysis. The free parameters in the 21-cm model are the redshift when the transition reaches 50% and the duration of reionization $\Delta z = (d\bar{x}_{HI}/dz)^{-1}\big|_{\bar{x}_{HI}=0.5}$. The model is sufficient to account for all of the visible features in the observed spectrum after spectral channels with radio-frequency interference have been masked from the data set.

We fitted the model to all available trials of $z_r$ in the observed spectrum using an approximately 20-MHz subset of the spectrum centred on $\nu_r = 1{,}420/(1 + z_r)$ MHz. In practice, we found that the order of polynomial that yields the best results is dependent on the trial redshift because the magnitude of the systematic structure in the spectrum varies with frequency. We used $m = 4$ for $z_r < 9$ and $m = 5$ for $z_r > 9$. By employing the prior that reionization was equally likely to have occurred at any redshift between $6 < z_r < 13$ and treating each frequency trial as an independent measurement, the observations exclude reionization histories shorter than $\Delta z < 0.19$ at 68% statistical confidence. Systematic uncertainty is estimated through inspection of the distribution function of best-fit derivatives, $\Delta z^{-1}$, from all frequency trials. Under a null hypothesis, we expect the distribution to peak at $\Delta z^{-1} = 0$ with large deviations indicative of systematic errors. We set the systematic error as the 68th percentile of the derivative distribution, corresponding to $\Delta z_{sys} = 0.21$. Combining statistical and systematic uncertainties in derivative space in quadrature yields our final confidence bounds of $\Delta z_{68} < 0.13$ at 68% combined confidence and $\Delta z_{95} < 0.06$ at 95% confidence. The excluded duration bounds as a function of reionization redshift $z_r$ are plotted in Fig. 2. These constraints are sufficient to rule out the most rapid plausible reionization histories, although more general theoretical expectations[3] currently yield predictions of $1 < \Delta z < 10$. Our result extends findings by WMAP, which ruled out at a fiducial instantaneous transition at $z < 7$ and yielded a best-fit $z_r = 10.5 \pm 1.2$.

The method demonstrated here is the only mechanism available at present to probe directly the derivative of the neutral fraction with redshift in the early Universe, and hence, to offer a unique way to constrain the reionization history. CMB anisotropy measurements probe reionization indirectly through an integral constraint on the optical depth to Thomson scattering of CMB photons and large-scale features in the E-mode polarization by free electrons in the intergalactic medium after reionization, whereas high-redshift quasar absorption spectra and Lyman-α galaxy surveys provide only a snapshot of the neutral fraction at the end of reionization.

Future enhancements to the EDGES instrument are forecast[11] to improve the constraints presented here by an order of magnitude and should be particularly valuable in combined analysis[12] with CMB and Lyman-α quasar spectra. The current measurement also serves as a pathfinder for extending the techniques to higher redshifts (lower frequencies) of $z \approx 20$ to search for absorption signatures[4,11,13] in the global 21-cm signal that should be more distinct than the reionization transition, but may be masked by larger foreground contributions. Such high-redshift global 21-cm observations may eventually provide unparalleled information about the ultraviolet emission and radiative feedback from the very first stars through the Wouthuysen–Field-effect-induced coupling of the kinetic and spin temperatures of the hydrogen gas in the intergalactic medium during the epoch of first light.

**Supplementary Information** is linked to the online version of the paper at www.nature.com/nature.

**Acknowledgements** This work was supported by the NSF. J.D.B was supported during part of this work at the California Institute of Technology by NASA through a Hubble Fellowship awarded by the Space Telescope Science Institute, which is operated by the Association of Universities for Research in Astronomy for NASA. This scientific work uses data obtained from the Murchison Radio-astronomy Observatory. We acknowledge the Wajarri Yamatji people as the traditional owners of the Observatory site. We thank CSIRO and Curtin University for logistical support and D. DeBoer, D. Herne, D. Emrich, M. Halleen and C. Halleen for on-site support.

**Author Contributions** Both authors contributed equally to the work in this paper.

**Author Information** Reprints and permissions information is available at www.nature.com/reprints. The authors declare no competing financial interests. Readers are welcome to comment on the online version of this article at www.nature.com/nature. Correspondence and requests for materials should be addressed to J.D.B. (judd.bowman@asu.edu).

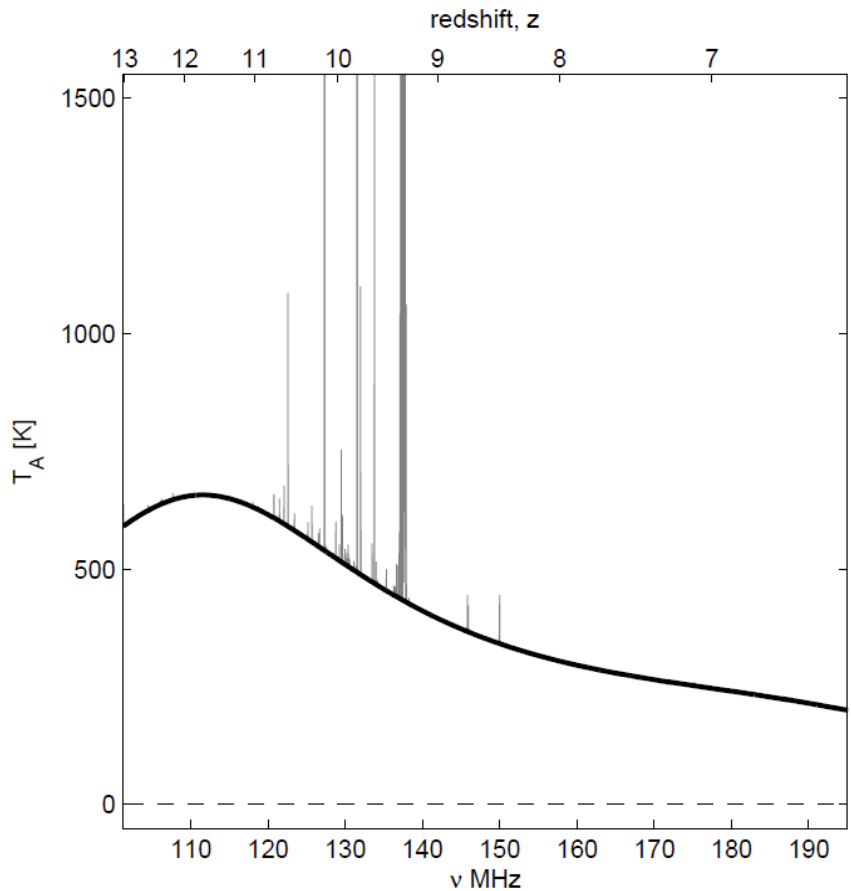

**Figure 1 Measured spectrum between 100 and 195 MHz.** The spectrum corresponds to redshifts $13 > z > 6$. The grey spikes are spectral channels that experienced radio-frequency interference during the integration and are masked from the analysis. The shape and amplitude of the spectrum are dominated by Galactic synchrotron emission and modulated by the uncalibrated antenna bandpass, which causes the spectrum to roll off from the characteristic $T_F \propto \nu^{-2.5}$ power-law form of the foregrounds at low and high frequencies. Any global 21-cm contribution in the spectrum is at the 20–30 mK level, approximately four orders of magnitude below the visible foreground emission. Thermal noise in the spectrum is 6 mK at 150 MHz using 1-MHz binned spectral resolution. The thermal noise increases at lower frequencies owing to the larger sky noise and lowered transmission efficiency of the antenna. Any 20-MHz sub-band in this spectrum can be fitted by a fifth-order polynomial, leaving residuals at or below the thermal noise level.

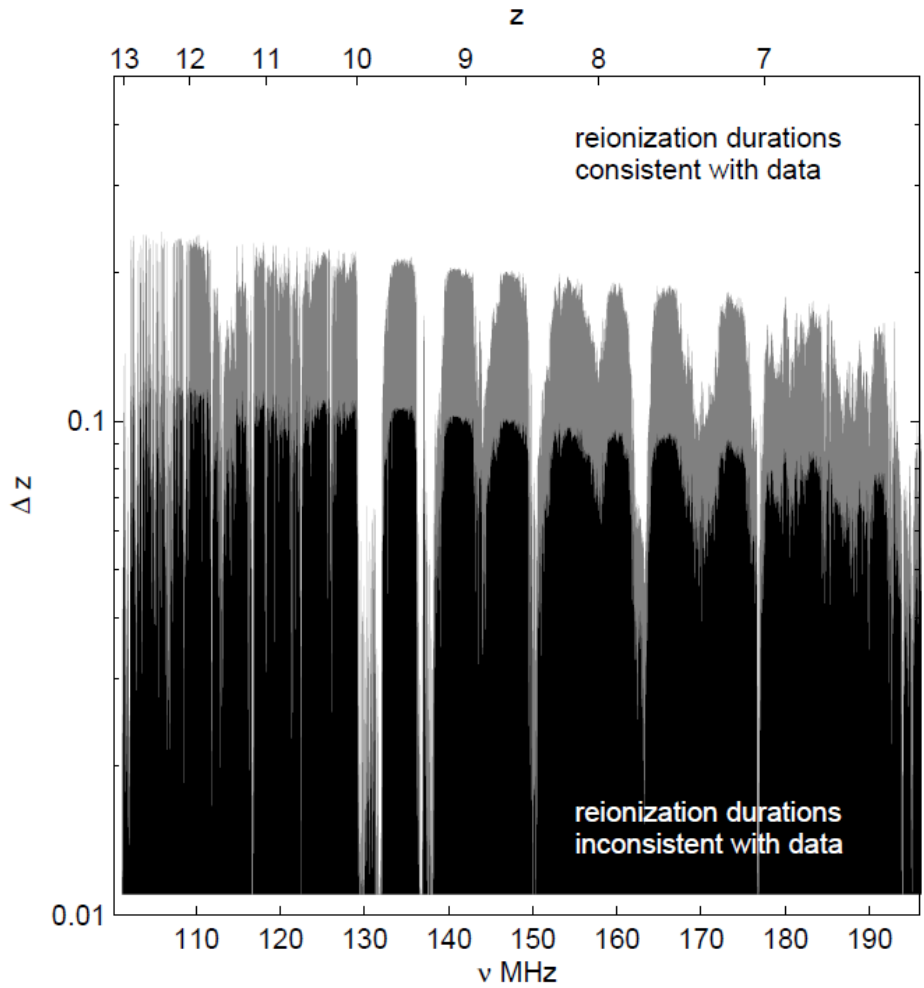

**Figure 2 Lower confidence bounds on the duration of the reionization transition. Statistical and systematic uncertainties are included.** Grey indicates the 68% confidence bound and black the 95% bound. The white region is allowed by the data. The data rule out rapid reionization histories shorter than $\Delta z \approx 0.1$ for many redshifts between $6 < z < 13$. The two large gaps at redshifts $z \approx 9.5$ (138 MHz) and $z \approx 10$ (130 MHz) are at frequencies that require extensive radio-frequency interference excision because they fall into satellite and aircraft communication bands, respectively.